\documentclass{article}

\usepackage{graphicx}
\usepackage{dcolumn}
\usepackage{bm}
\usepackage{graphics}
\usepackage{graphicx}
\usepackage{amsmath,amssymb}
\usepackage{color}

\title{Maintaining extensivity in evolutionary multiplex networks}
\author{Chris G. Antonopoulos\\Department of Mathematical Sciences, University of Essex,\\Wivenhoe Park, UK \\\and Murilo S. Baptista\\Institute of Complex Sciences and Mathematical Biology,\\University of Aberdeen, SUPA,\\Aberdeen, UK\\}

\date{\today}
\begin{document}

\maketitle

\begin{abstract}
In this paper, we explore the role of network topology on maintaining the extensive property of entropy. We study analytically and numerically how the topology contributes to maintaining extensivity of entropy in multiplex networks, i.e. networks of subnetworks (layers), by means of the sum of the positive Lyapunov exponents, $H_{KS}$, a quantity related to entropy. We show that extensivity relies not only on the interplay between the coupling strengths of the dynamics associated to the intra (short-range) and inter (long-range) interactions, but also on the sum of the intra-degrees of the nodes of the layers. For the analytically treated networks of size $N$, among several other results, we show that if the sum of the intra-degrees (and the sum of inter-degrees) scales as $N^{\theta+1},\;\theta>0$, extensivity can be maintained if the intra-coupling (and the inter-coupling) strength scales as $N^{-\theta}$, when evolution is driven by the maximisation of $H_{KS}$. We then verify our analytical results by performing numerical simulations in multiplex networks formed by electrically and chemically coupled neurons.
\end{abstract}

\section{Introduction}\label{Intro}
Complex networks are ubiquitous in nature, studied in Physics and other disciplines. They are composed of different components, which are connected between them in a non-trivial way. They range from biological networks, such as neural to technological networks, to the internet, power-grids and transportation, to social and affiliation networks, etc. \cite{Newman2010}. A significant point of interest is to understand the action of the type of coupling in the behaviour of the complex network. For example, all-to-all coupled dissipative systems can exhibit collective chaotic behaviour with macroscopic variables showing irregular behaviour due to the interplay of correlations among the different components of the system. These noticeable properties make them less well understood than systems with short-range interactions \cite{Takeuchietal2011}.

A key factor is the role of chaos, but equally important is the way the various components of the system are connected between them, by short- or long-range interactions, or all-to-all connections. It is well-accepted that systems with short-range interactions are characterised by extensive chaotic behaviour, as conjectured in \cite{Ruelle1982}, namely by quantities that grow linearly with the increase of the system size $N$. The main argument is that a sufficiently large spatial domain can be divided into small, independent subsystems with similar dynamical properties.

A well-established method of characterising chaotic behaviour is the calculation of Lyapunov exponents (LEs), which quantify the rate of divergence of infinitesimally close trajectories in phase-space \cite{Skokos2010}. As conjectured by Ruelle in 1982 \cite{Ruelle1982}, extensivity can be inferred by studying the scaling properties of the curve of the LEs, $\lambda_i$, after they are ordered in descending order (i.e. Lyapunov spectrum) and are plotted as a function of their normalised index $i/N$, with respect to the system size $N$. If the curve of the Lyapunov spectrum (for a normalised index) collapses onto a single asymptotic curve as $N$ grows, then one talks about extensive chaos. This is indeed true for several examples of systems, such as large, spatially extended dissipative systems, for generic models of spatiotemporal chaos in one dimensional spaces \cite{Takeuchietal2009}, for locally coupled systems \cite{Pauletal2007}, and for dissipative systems in a random network topology with sparse connectivity \cite{Lucciolietal2012,monteforte2010dynamical}.

In contrast to what was found in \cite{Takeuchietal2009,Pauletal2007}, in higher dimensional spaces or for all-to-all coupled systems, extensivity as a measure of how a network variable or behaviour scales linearly with the size of the network, is questioned. It is well-known that for highly dimensional chaotic systems, nontrivial collective behaviour (e.g. the mean field, or any other field) is observed to evolve periodically, quasi-periodically, or even chaotically in time  \cite{Takeuchietal2009,Chateetal1992,Kaneko1990,Pikovskyetal1994,Lucciolietal2012}. Since, collective modes are by definition intensive quantities, i.e. they do not depend on the system size, it would imply that systems with non-trivial behaviour can not be extensive \cite{Takeuchietal2009}, even though the dynamics of the network is chaotic. To date, it is still elusive the general scenario of how extensivity appears in complex networks. According to \cite{Takeuchietal2011}, for identical units submitted to the same self-consistent forcing, the influence of a given unit on the mean field vanishes in the thermodynamic limit, suggesting that LEs should become equal, what points to extensive behaviour. These results are in contrast to the evidence provided by the same authors for LEs of finite-size globally all-to-all coupled systems that are seen to become unequal, implying non-extensive chaotic behaviour. A step forward in this subject was provided by the work in \cite{Lucciolietal2012}, where it is shown that the in-degree (i.e. the number of incoming connections per node) in three classes of random networks with all nodes possessing equal in-degrees plays an important role to the extensive character of sparsely and densely connected networks, when the strength of the connection is constant.

In this paper, we explore the role of network topology on maintaining the extensive property of entropy. We derive and study an analytical formula for the sum of positive LEs, $H_{KS}$ (see Eq. \eqref{final_HKS_1_network}), a quantity closely related to the entropy in bounded deterministic systems \cite{Pesin1976,Pesin1977,Eckmann1985,Ruelle1978}, from which one can clearly see the network characteristic conditions for $H_{KS}$ to become either an extensive, sub- or super-extensive thermodynamic quantity, assuming chaotic behaviour in the complex network. We consider networks evolving from an initial multiplex \cite{arenasprl,MartinHernandezetal2014,PhysRevE.88.032807,PhysRevX.3.041022,Boccaletti20141,kivela2014multilayer} configuration, formed by two layers of nodes connected internally by short-range interactions, to a final network, characterised by a mixture of short- and long-range interactions interconnecting the two layers. We show that extensivity relies not only, as previously reported in \cite{Takeuchietal2009,Chateetal1992,Kaneko1990,Pikovskyetal1994}, on the interplay between the coupling strengths of the dynamics associated to the short- and long-range interactions or on the in-degree \cite{Lucciolietal2012,monteforte2010dynamical}, but also on quantities never before associated to extensive behaviour: On the sum of the intra-degrees of the nodes of the layers, or equivalently on the sum of the eigenvalues of the connecting Laplacian matrices of the layers, and also on the sum of inter-degrees, or equivalently on the total number of inter-connections. A consequence of our theoretical results is that for networks whose sum of intra- and sum of inter-degrees both scale as $N^{\theta+1},\;\theta>0$, extensivity can be maintained if the intra- and inter-coupling strengths scale as $N^{-\theta}$. This scaling was obtained by evolving the network by a maximisation process for $H_{KS}$, requiring that as the network grows in size, it preserves the positiveness of all LEs.
 
We then provide numerical evidence that shows that to maintain the extensive character of entropy in a multiplex network of coupled Hindmarsh-Rose (HR) neurons \cite{Hindmarshetal1984}, one needs to evolve the network by a process that keeps the intra-coupling (electrical) strengths constant whereas the sum of the intra-degrees of the nodes scales linearly with the number of nodes of the network, a constrain that is expected from our analytical results from the multiplex networks of discrete maps.  We consider several initial multiplex network configurations with different sizes composed of {\it non-equal} layers of neurons connected by short-range connections with small-world intra-topologies \cite{Wattsetal1998}, a structure inspired by the anatomical networks found in the human brain \cite{He01102007}. For each multiplex network, the evolutionary process adds new, long-range inter-connections between the layers and finds appropriate coupling strengths to maximise $H_{KS}$, maintaining the small-world topology.

Finally, we show for which network characteristics one should expect to find that extensivity in $H_{KS}$ implies the existence of an invariant curve for the ranked LEs (namely, for the Lyapunov spectra). Whereas for single networks extensivity typically implies the existence of an invariant curve for the Lyapunov spectra, in multiplex networks this correspondence can only exist if certain network characteristics are maintained during network evolution. This result is relevant since it shows that previous results from the literature (e.g., \cite{Ruelle1982}, \cite{politi1986}) for single networks do not necessarily apply to multiplex networks.
     
\section{Materials and Methods}

\subsection{The Calculation of Lyapunov Exponents}

Given a multiplex network $G$ formed by coupled maps with dimensionality $\mathbb{R}^N$ with a constant Jacobian 
\begin{equation}
\mathbf{J} = C \mathbb{I} - \mathbf{L}, 
\label{methods-jacobian}
\end{equation}
where $\mathbf{L}$ represents a Laplacian matrix, $C$ a constant and $\mathbb{I}$ the identify matrix, the LEs are given by
\begin{equation}
\mathbf{\lambda} = \log{|| \mathbf{J} ||},  
\label{methods-LE-vector-map} 
\end{equation}
\noindent
where $||.||$ denotes the absolute values of the eigenvalues $\mu_i$ of the argument, which therefore implies that each LE can be calculated by 
\begin{equation}
\lambda_i = \log{|C - \mu_i|},  
\label{methods-LEi-map} 
\end{equation}  
\noindent
where $\mu_i$ represents the eigenvalue $i$ of $\mathbf{L}$.

There is an alternative way to calculate the LEs of a map with a constant Jacobian if it possesses a synchronisation manifold. As shown in \cite{baptista2011complex,baptista_arxiv2015}, due to the fact that the synchronisation manifold exists and the dynamics in $G$ has a constant Jacobian, the LEs of the synchronisation manifold and its transversal directions (whose values can be calculated analytically) are equal to the LEs of the attractors appearing in the dynamics of the multiplex network. Our interest is in the calculation of the LEs of these attractors.

For a linear system of differential equations in $\mathbb{R}^N$ \cite{barreira2002} such as 
\begin{equation}\label{methods-dynamical_system_diffusive_process}
\dot{\vec{x}}=C\vec{x}-\mathbf{L}\vec{x},
\end{equation}
with the same Jacobian as in (\ref{methods-jacobian}), the LEs can be calculated by 
\begin{equation}\label{methods-lyapunov-exponent-flows}
\lambda_i = \log{(e^{\mathbf{C - \mathbb{\mu}_i}})},  
\end{equation}
\noindent
and therefore for a linear system of differential equations, the LEs are the eigenvalues of $\mathbf{J}$.

For a system of nonlinear differential equations, the spectrum of 
LEs is calculated in this work using the numerical method in Ref. \cite{wolf1985}.

\subsection{Network models considered}

We consider initially two non-connected layers $G_1=({N}_1,{E}_1)$ and $G_2=({N}_2,{E}_2)$, where $({N}_1,{N}_2)$ is the set of nodes and $({E}_1,{E}_2)$ the set of edges. For simplicity, we suppose throughout the paper, that all networks considered are undirected, i.e., each connection between two nodes is bidirectional. The network resulting from these two non-connected layers is a new network $({N},{E})$, where $N$, $E$ are the number of nodes and edges respectively. For the latter network, we then define $l_{12}$ to be the number of undirected inter-connections between the two layers $G_1$ and $G_2$ with $l_{12}\leq N_1^2$, where $N_1$ is the number of nodes in $G_1$. We finally define $G$ to be the multiplex network formed by the two layers and their $l_{12}$ inter-connections.

Following \cite{baptista_arxiv2015}, our first network model we consider is a multiplex network $G$ 
with dynamics for its nodes described by the shift map
\begin{equation}\label{coupled_shift_map}
\vec{x}_{n+1}=2\vec{x}_n-\mathbf{L} \vec{x}_n,\mod 1
\end{equation}
(discrete multiplex network), where $n$ is the iterations subscript, $\mathbf{L}=\epsilon \mathbf{L^B} + \gamma\alpha \mathbf{L^A}$ with $\epsilon$ being the coupling strength of the intra-connections in $G_1$ and $G_2$, $\gamma$ the coupling strength of the inter-connections between $G_1$ and $G_2$, and $\alpha=l_{12}/N_1$.
$\mathbf{L^B} = \left( 
\begin{array}{cc}
\mathbf{B}  &  0  \\
0  & \mathbf{B}  
\end{array}
\right) $ is the Laplacian of the intra-connections and $\mathbf{L^A} = \left( 
\begin{array}{cc}
  \mathbf{D}_1 &  -\mathbf{A}   \\
 -\mathbf{A}^T   & \mathbf{D}_2  
\end{array}
\right)$ the Laplacian of the inter-connections, where $T$ stands for the transpose. We consider two identical layers $G_1$ and $G_2$, connected by $l_{12}$ inter-connections. Each node in $G_1$ makes an equal number of inter-connections to a corresponding node in $G_2$.

$\mathbf{L^B}$ represents the Laplacian of the two uncoupled complex networks and their intra-connections (the Laplacian $\mathbf{B}$) and $\mathbf{L^A}$ the Laplacian of the inter-connections between the layers. $\mathbf{D}_1$ and $\mathbf{D}_2$ represent the identity degree matrices of the adjacency matrices $\mathbf{A}$ and $\mathbf{A}^T$, respectively. Their components are defined as $(\mathbf{D}_1)_{ii}=\sum_j {A}_{ij}$ and $(\mathbf{D}_2)_{ii}=\sum_j {A}^T_{ij}$, both with null off-diagonal terms.

We will also consider continuous multiplex networks of Hindmarsh-Rose (HR) neurons, $G$, whose equations of motion are given by 
\begin{flalign}\label{HR_model_Nneurons}
 \dot{p}_i&=q_i-a p_i^3+bp_i^2-n_i+I_{ext}-\epsilon\sum_{j=1}^{N}{\bf L^{A}}_{ij}H(p_j)-\gamma(p_i-V_{syn})\sum_{j=1}^{N}{\mathbf{A}}_{ij}K(p_j)\nonumber,\\
 \dot{q}_i&=c-dp_i^2-q_i\nonumber,\\
 \dot{n}_i&=r[s(p_i-p_0)-n_i],\;i=1,\ldots,N,
\end{flalign}
where $H(p_i)=p_i$ and $K(p_j)=\frac{1}{1+e^{-\lambda(p_j-\theta_{syn})}}$ \cite{Baptistaetal2010}. 
We use $a=1$, $b=3$, $c=1$, $d=5$, $s=4$, $p_0=-1.6$, $r=0.005$ and $I_{ext}=3.25$. For these values, each neuron can exhibit chaotic behaviour and the solution for $p$ exhibits typical multi-scale chaotic behaviour characterised by spiking and bursting activity consistent with the membrane potential observed in experiments on single neurons \textit{in vitro} \cite{Hindmarshetal1984}. Thus, chaos not only allows the reproduction of behaviours empirically observed in experiments with single neurons, but also allows the networks to process information \cite{Baptistaetal2008A}. We also set $\theta_{syn}=-0.25$, $\lambda=10$ and $V_{syn}=2$ to create excitatory post-synaptic couplings. In Eqs. \eqref{HR_model_Nneurons}, $\gamma$ is the coupling strength associated to the chemical inter-connections between neurons of the two layers and $\epsilon$ to the electrical intra-connections between neurons within $G_1$ and $G_2$. $\mathbf{L^{A}}$ accounts for the way neurons are electrically (diffusively) coupled and it is a Laplacian matrix as defined earlier and $\mathbf{A}$ represents the adjacency matrix of the inter-connections.

\section{Results}

\subsection{Extensivity in the discrete multiplex networks with only positive LEs}
 
From \cite{baptista_arxiv2015} and from Eq. (\ref{methods-LE-vector-map}), the LEs of the discrete dynamical system in Eq. \eqref{coupled_shift_map} are given by
\begin{equation}\label{LEs_coupled_shift_map}
\lambda_{i}=\log|2-\mu_i|,\;i=1,\ldots,N,
\end{equation}
where $\mu_{i}$ are the eigenvalues of $\mathbf{L}$. 

Notice also that since map (\ref{coupled_shift_map}) has a constant Jacobian, the quantities calculated in this work are not affected by the mod function, and therefore, manipulating the trajectory to fall into $[0,1]$, what is typically carried out in the study of map lattices, is not necessary here, since the locations of the points of the trajectory do not affect the analytical derivations.

From the LEs, we can calculate
\begin{equation}\label{HKS_entropy}
H_{KS}  = \sum_{\lambda_i>0}\lambda_i.  
\end{equation}

Following \cite{MartinHernandezetal2014}, $\mu_{i}$ are related to the eigenvalues $\omega_i$ of the Laplacian $\mathbf{B}$ of the subnetwork by
\begin{flalign}\label{final_eigenvalues}
\mu_{2i-1}&=\epsilon\omega_{i},\nonumber\\
\mu_{2i}&=\epsilon\omega_{i} + 2\gamma\alpha, 
\end{flalign}
where $i=1,\ldots,N_1$.

For a given initial multiplex network $G^{0}$ with $N^0$ nodes and $\tilde{N}^{0}$ positive LEs (where $N^0\geq\tilde{N}^{0}$), one way to maintain extensivity in $H_{KS}$ is by keeping constant during evolution the ratio $N^0/\tilde{N}^0$. This necessary requirement is a constrain imposed to the network and maintains the nature of the Lyapunov spectra. As this network then grows to a network $G^1$ with $N^1$ nodes (with $N^1>N^0$), we require that $N^0/\tilde{N}^0=N^1/\tilde{N}^1$, where $\tilde{N}^1$ is the number of positive LEs for network $G^1$. 

Our first analysis will be done considering networks that have only positive LEs, a process leading to the maximisation of $H_{KS}$. This network has nodes that are fully de-coherent and asynchronous, even thought the synchronisation manifold exists. Thus, a natural choice is to set $N^0 = \tilde{N}^0$ and $N^1 = \tilde{N}^1$, leading to networks that maintain all LEs positive. Consequently, as the network grows, $\epsilon$ must decrease accordingly so that all LEs of the network are positive. This choice also allows us to expand $\log\bigl|1-\epsilon\frac{\omega_i}{2}\bigr|$ and $\log\bigl|1-\epsilon\frac{\omega_i}{2} - \gamma\alpha\bigr|$ in Taylor expansions, keeping only up to the first order terms, to obtain Eqs. \eqref{HKS-single} and \eqref{final_HKS_1_network}.

To develop understanding about how topology promotes extensivity, sub- or super-extensivity, we start by analysing an isolated network given by one of the two layers, meaning that $\gamma$=0 and $N=N_1$. In the following, we notice that \cite{Chung1997}
\begin{equation}\nonumber
\sum_{i=1}^{N_1}\omega_i=\sum_{i=1}^{N_1}d_i\equiv S,
\end{equation}
where $d_i$ is the intra-degree of node $i$ in $G_1$ or $G_2$.
In such a case, 
\begin{equation}
H_{KS}\approx N_1\log{(2)}-\frac{\epsilon}{2}S,
\label{HKS-single}
\end{equation}
by combining Eqs. \eqref{LEs_coupled_shift_map}, \eqref{HKS_entropy} and \eqref{final_eigenvalues}. In an all-to-all network, $S \propto N_1^2$ and extensivity in $H_{KS}$ can be maintained if  and $\epsilon$ is rescaled by $\epsilon \propto \frac{1}{N_1}$. This rescaling is obtained by imposing a number of $N_1$ positive LEs (which in turn prevents the onset of full synchronisation), which can be achieved for $\epsilon > \frac{1}{\omega_{N_1}} = \frac{1}{N_1}$, according to Eq. \eqref{LEs_coupled_shift_map}. 

Let us consider a circulant graph $G$ now, where every node is connected to $k$ nodes in a regular way. In this case, $S = kN_1 \propto N_1$, which can be seen as a rough model of a small-world network. In order to maintain extensivity in $H_{KS}$, $\epsilon$ must be kept constant. Moreover, we also require that $\epsilon$ be sufficiently small for the Taylor expansions to be valid. Since the largest eigenvalues of this network scale as $2k$ \cite{van2010graph}, and assuming that $k$ is sufficiently large, to maintain all LEs positive, we require that $\epsilon > \frac{1}{\omega_{N_1}}$, or $\epsilon > \frac{1}{2k}$, regardless of the value of $N_1$. 

The choice of $\epsilon \propto N_1$ would lead to a super-extensive $H_{KS}$. In this case, the number of positive LEs is equal to $N_1$, but their intensity is enlarged. Generalising, if we now assume that $S \propto N_1^{\theta+1},\;\theta>0$, extensivity can be maintained if $\epsilon  \propto N_1^{-\theta}$ (or, in the thermodynamic limit, if $\epsilon  \propto N_1^{-\theta-1}$, so $\epsilon S$ is constant and therefore, $\lim_{N \rightarrow \infty} H_{KS}/N$ is a constant). Thus, supposing that the growing process imposes that $S(N_1) \propto N_1^{\theta+1}$, extensivity can be maintained by requiring that $\epsilon(N_1) \propto N_1^{-\theta}$. A surprising consequence of our results is that the topology of a network is not the most important factor to determine extensivity, but rather $S$, i.e. the sum of intra-degrees. For example, both regular graphs and Erd\H{o}s-R\'enyi random graphs whose $S$ scales as $N$, can provide extensive networks if the intra-coupling strengths are made constant. As a consequence, if random networks whose nodes have equal degrees and constant coupling strengths are considered, as in \cite{Lucciolietal2012}, our work suggests that extensivity could appear in these networks if the coupling strength is inversely proportional to the in-degree. Taking as an example the numerical results shown in Fig 1 of \cite{Lucciolietal2012}, extensivity is found when the coupling strength $g=0.1$ and for in-degree satisfying $K>60$, which agrees with our predictions since they would suggest that extensivity is found for $K>\beta \frac{1}{0.1}$, where $\beta$ represents a proportionality constant.

We now study how topology is related to extensivity in multiplex networks. Combining Eqs. \eqref{LEs_coupled_shift_map}, \eqref{HKS_entropy} and \eqref{final_eigenvalues}, we get
\begin{equation}\label{final_HKS_1_network}
H_{KS} \approxeq  N\log{(2)} -\frac{\epsilon}{2} S^{\prime} - \gamma l_{12},
\end{equation}
where $ S^{\prime}=2S$ and represents the sum of the degrees of all nodes in the multiplex network. This makes Eq. \eqref{final_HKS_1_network} an implicit function of $N$ that can capture better the characteristics of our simulations on the evolving neural networks.

The previous analysis made for single networks that resulted in Eq. \eqref{HKS-single} remains valid for multiplex networks (see Eq. \eqref{final_HKS_1_network}), with the additional contribution from the inter-couplings, given by $l_{12}$ in Eq. \eqref{final_HKS_1_network}. If $\epsilon$ is set to maintain extensivity (i.e. $\epsilon \propto \frac{N}{S^{\prime}}$) and the network is growing under the constrain that  $l_{12}\propto\zeta N_1\propto\frac{\zeta N}{2}$, with $\zeta\propto N^{\theta}$ (noticing that $\zeta<N_1$) and thus $l_{12} \propto N^{\theta+1}$, then extensivity in $H_{KS}$ can be maintained if $\gamma \zeta$ is constant, leading to $\gamma \propto N^{-\theta}$. Sub-extensivity occurs if $\gamma \zeta \propto N^{\theta}$ with $\theta<0$ and super-extensivity if $\gamma \zeta \propto N^{\theta}$ with $\theta>0$. However, $H_{KS}$ remains super- or sub-extensive for only a finite and small number of evolution iterations, since the eigenvalues in Eq. (\ref{final_eigenvalues}) are only valid for multiplex network-configurations for which the total number of inter-connections is smaller than $N_1^2$, since $l_{12}<N_1^2$, resulting therefore to $1<N^{\theta}<N_1$.

\subsection{Extensivity in the discrete multiplex networks with positive and negative LEs}

We now consider the case in which the multiplex network of Eq. (\ref{coupled_shift_map}) has positive as well as negative LEs. In particular, we study the case where there is a number $u$ of negative exponents and that the inter-coupling strength $\gamma$ is responsible for the change in the sign of LEs, i.e., $\log{|2 - \epsilon \omega_i |} >0$ and $\log{|2 - \epsilon \omega_i -2 \gamma \alpha|} > 0$ ($\log{|2 - \epsilon \omega_i -2 \gamma \alpha|} \leq 0$) for $i \leq N_1 - u$ (for $i > N_1 - u$). We also assume as before that $\epsilon \omega_i$ is small so that we can expand $\log{(1-\epsilon \omega_i)} \cong - \epsilon \omega$, as done previously in our derivations. However, now we need to assume that the inter-coupling strength is not arbitrarily small in order to produce negative LEs. Therefore, we use the expansion 
$\log{(1-\frac{\epsilon \omega_i}{2} - \gamma \alpha)} \cong \log(1-\gamma \alpha) - \frac{\epsilon \omega_i}{2(1-\gamma \alpha)}$. We obtain that

\begin{equation}\label{final_HKS_1_network-negativeLE}
H_{KS} \approxeq  (N-u) \log{(2)} - \frac{\epsilon}{2}S - \frac{\epsilon}{2(1-\gamma \alpha)} \sum_{i=1}^{N_1-u} \omega_i + (N_1 - u)  \log{(1-\gamma \alpha)},
\end{equation}
\noindent assuming $u$ is constant. The conclusion is that if the network is evolved by preserving the number $u$ of negative LEs, then previous arguments for extensivity extend to this type of network if $\sum_{i=1}^{N_1-u} \omega_i \propto N$. If $u$ is not constant, extensivity can be maintained by evolving the network in such a way to make $u$ a linear function of $N$, and therefore, the number of negative LEs must scale linearly with the size of the multiplex network. The last right-hand term of Eq. (\ref{final_HKS_1_network-negativeLE}) would be given by $\sum_{i=1}^{N_1 - u} \log{(1-\gamma \alpha)}$ for a non-constant $u$.

\subsection{Extensivity in the continuous HR multiplex network}

In our analytical derivations from Sec.``Extensivity in the discrete multiplex networks", we can control the number of positive LEs of the discrete multiplex network. The same though cannot be done for our numerical analysis of the multiplex network  of HR neurons. However, in our further numerical simulations, we search for electrical and chemical coupling parameters in the HR system that maximise the sum of positive LEs.

\begin{figure}[!ht]
\centering{
\includegraphics[scale=0.42]{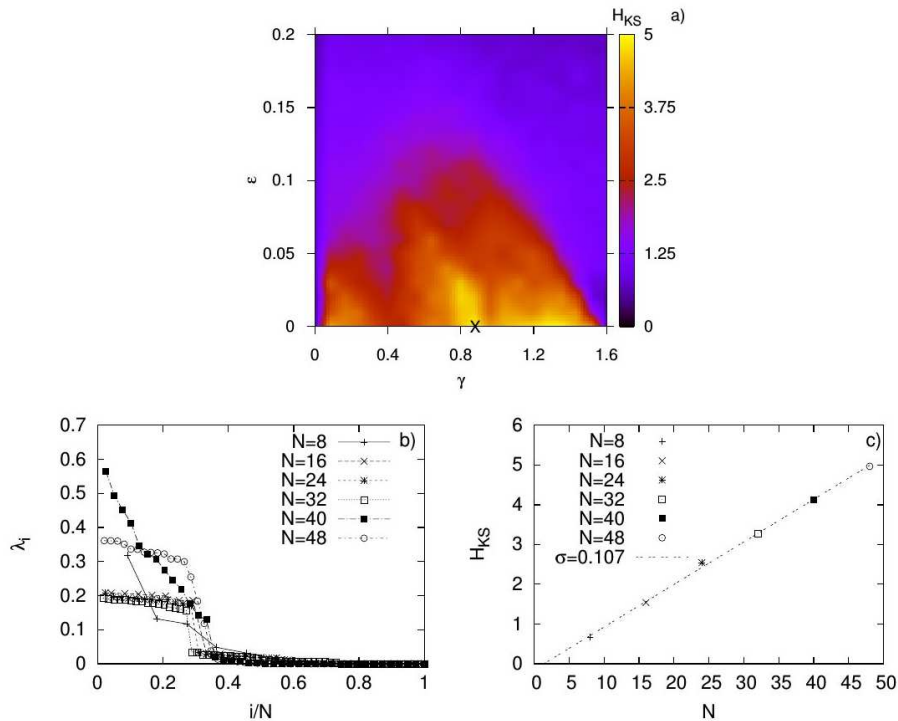}
\caption{\textbf{Extensivity in neural network evolution by maximising $\pmb{H_{KS}}$}. Panel a): An example of the parameter space of chemical $\gamma$ and electrical coupling $\epsilon$ for $N=48$ neurons arranged in two equally-sized small-world layers. The \textbf{X} point corresponds to the coupling pair for which $H_{KS}$ is maximal in the parameter space. Panel b): Plot of the Lyapunov spectra for different network sizes $N$ and Panel c): The linear relation between $H_{KS}$ and network size $N$, where $\sigma$ is the slope of the linear fitting to the data.}\label{results_maximize_Hks_neural_network_evolving_technique_H_KS}}
\end{figure}

To check whether a neural network that evolves by a process that maximises $H_{KS}$ remains extensive, we performed detailed numerical simulations considering two evolving interconnected small-world \cite{Wattsetal1998} layers $G_1$ and $G_2$ with Hindmarsh-Rose (HR) dynamics \cite{Hindmarshetal1984} for their nodes. During evolution, we add, in a random fashion, and retain in $G$ new inter-connections if they lead to an increase in $H_{KS}$ compared to its value before the addition. 

We evolve the dynamics of Eq. \eqref{HR_model_Nneurons} and calculate all LEs $\lambda_i,\;i=1,\ldots,(3N)$ \cite{Benettin1980a,Benettin1980b} sorted in descending order. In more details, following \cite{Antonopoulosetal2015}, we consider a starting multiplex network $G$ composed of two layers $G_1$ and $G_2$, connected initially with a single chemical inter-connection. Nodes in both layers are only electrically connected. Each subnetwork is equipped with a small-world structure \cite{Wattsetal1998}, having although different internal connectivities. For us to be able to compare our numerical results with the analytical one in Eq. \eqref{final_HKS_1_network}, we consider that both layers have the same number of nodes. However, to place realism in our simulations, we construct them so that they have different small-world structures. Having a network with a given total number of nodes $N$, we fix the values of the inter- and intra-coupling strengths, and then evolve the starting network $G$ by adding new chemical excitatory inter-connections linking a node in $G_1$ to a node in $G_2$. The electrical connections in both layers, their topologies and, values of $\gamma$ and $\epsilon$ are kept fixed and independent of $N$ during evolution. We only increase the number of inter-connections and thus $\alpha$. The criterion to whether add a new chemical inter-connection is based on whether the newly added connection will lead to an increase in $H_{KS}$ prior to the addition of the inter-connection. If an added inter-connection does not contribute to an increase in $H_{KS}$, then the new edge is deleted in $G$, and the search for another one starts in an iterative manner, terminating when the maximum number of possible pairs of nodes is reached. The pairs of nodes in $G_1$ and $G_2$ are chosen randomly, and are equiprobable. We repeat this evolution process considering the same initial network, for different pairs of coupling strengths $\epsilon$ and $\gamma$. We then evolve several such initial configurations, i.e. a network with fixed $N$ for several coupling strengths and pick the network whose final evolution, starting with size $N$ and for the specified pair of coupling strengths, renders $H_{KS}$ maximal.

\begin{figure}[!ht]
\centering{
\includegraphics[height=10cm,width=7cm,angle=-90]{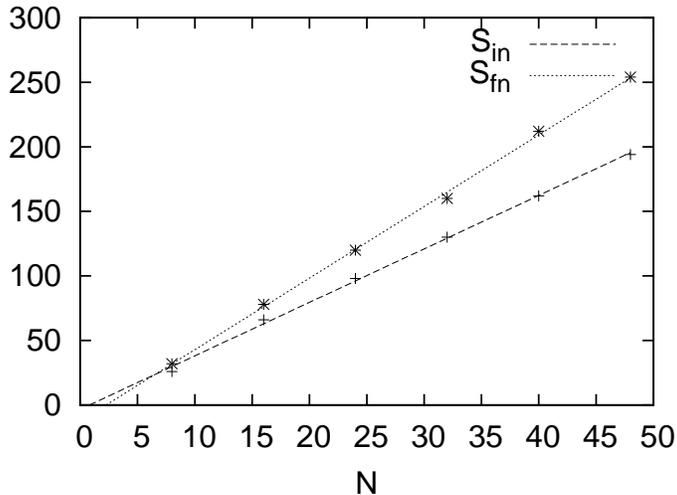}
\caption{\textbf{Sums of the degrees of the multiplex network grows linearly with the network size}. $S_{in}$ is the sum of the degrees of the adjacency matrices of the starting networks and $S_{fn}$ the sum of the degrees of the adjacency matrices of the finally evolved multiplex networks of the evolution process.}\label{results_sum_of_degrees_vs_N_plot_2_swc_inter_cc}}
\end{figure}

We present the results of this study in Fig \ref{results_maximize_Hks_neural_network_evolving_technique_H_KS}. The first panel shows an example of the parameter space $(\gamma,\epsilon)$ for $N=48$ neurons, which are equally arranged in the two small-world layers, $G_1$ and $G_2$. Point \textbf{X} denotes the coupling pair for which $H_{KS}$ becomes maximal in this space, after the end of the evolution process. For similar pairs of couplings and for six multiplex networks of increasing size $N$, we then computed their Lyapunov spectra at the end of the evolution processes, and present them in the second panel of the figure. The second and many of the larger LEs are seen to be positive, pointing to chaotic behaviour in the considered complex networks. It is also evident that all Lyapunov spectra start increasing from zero for about the same normalised index $i/N$. Using the values of these LEs we have then computed the corresponding $H_{KS}$ values presented in the third panel of the figure. We note that, irrespectively of the network size $N$ in our simulations, we found out that $H_{KS}$ becomes maximal for a roughly constant electrical coupling strength, as expected by our analytical results, since the initial network topologies have sum of degrees that scale linearly with $N$ (see Fig \ref{results_sum_of_degrees_vs_N_plot_2_swc_inter_cc}). It is also worth noting that $\epsilon\ll\gamma$, a result that points out the crucial role played by the nonlinear chemical inter-connections as opposed to the linear (electrical) ones for the maximisation of $H_{KS}$. The chemical connections have a crucial role in the increase of chaos and consequently on the network entropy. The third panel shows the extensive character of $H_{KS}$, scaling linearly with $N$, with a slope $\sigma\approx0.107\pm0.002$ obtained from the linear fitting to the data. Our results indicate that as long as the layers have node-degrees that sum up linearly with the size of the network, the evolutionary process finds the appropriate inter-connections to maintain the extensive character of $H_{KS}$.

The extensive character of the neural networks implies that the sum of the degrees, $S$, of the multiplex network $G$ grows linearly with $N$, initially as well as after the end of the evolution process. Indeed, this is what is happening and we present in Fig \ref{results_sum_of_degrees_vs_N_plot_2_swc_inter_cc}, where we show that the sum of the degrees of the initial $S_{in}$ and finally $S_{fn}$ evolved multiplex networks grow linearly with the system size $N$. According to Eq. (\ref{final_HKS_1_network}), if $S^{\prime}$ and $N_1l_{12}$ scale linearly with $N$, then the sum of the node-degrees, $S^{\prime}+N_1l_{12}$, should also scale linearly with $N$.

Panel b) of Fig \ref{results_maximize_Hks_neural_network_evolving_technique_H_KS} shows that the Lyapunov spectra do not approach an invariant curve. From our results considering the networks of maps, we have obtained that in order to maintain extensivity and to have invariant Lyapunov spectra, it is required that the average intra-degree $\bar{b}(N_1)$, the ratio $\frac{\bar{d}(N_1)}{\omega_{N_1}(N_1)}$ and the largest eigenvalue of the subnetworks, $\omega_{N_1}(N_1)$ remain all invariant during the evolution process. As shown in S1 file, these quantities do remain roughly invariant, however, have weak dependence on $N$, which could explain the non-invariance of the curve of the Lyapunov spectrum. Another factor that can contribute to the non-convergence of the curve of the Lyapunov spectrum is the finite nature of the studied networks. The study of even larger multiplex networks would allow one to point the source for the apparent non-invariance of the Lyapunov spectra curve.

Our analytical derivation for $H_{KS}$ in Eq. (\ref{final_HKS_1_network}) are based on the discrete multiplex network of coupled maps in Eq. (\ref{coupled_shift_map}), whereas the analysis for the HR multiplex neural network are based on numerical calculations. The correspondence between analytical (from the discrete multiplex network) and numerical results (from the continuous multiplex network) lays in that in both analyses we obtain that networks can grow by maintaining its extensivity character and that a relevant criteria to obtain extensivity is that the sum of the intra-degree of the layers of the multiplex network must be a linear function of the size of the network.

\subsection{Extensivity and the curve of Lyapunov exponents}

In Ref. \cite{Ruelle1982} it has been shown that an essential signature of extensive chaos is that the curve of the LEs, $\lambda_i$, after they are ordered in descending order (i.e. Lyapunov spectrum) and are plotted as a function of their normalised index $i/N$, with respect to the system size $N$, collapses to a universal curve as the size of the system $N$ increases. This property has been verified in all extensive systems observed so far as well as in spatially extended systems with diffusive coupling \cite{politi1986}. This approach is appropriate when an equation for the LEs as a function of ordered numbers such as the eigenvalues of the Jacobian $A$ of the dynamics is unknown. However, if such an equation is known, as in this work, we can instead consider a more natural measure, the probability density $\rho(\omega_j)$ of the eigenvalues of $A$. Besides, working with probability densities allows us to access the properties of the invariance of the Lyapunov spectrum based on space averages, which are easier to be tackled analytically. We will demonstrate in the following (more details can be found in S1 file) that even though extensivity can also be typically observed in our studied multiplex networks, depending on how topology and coupling strengths are altered as the network grows, the existence of an invariant Lyapunov spectrum may not be observed. On the contrary, and supporting previous works \cite{politi1986}, extensivity of a  single network, represented by the sum of the positive LEs, typically implies that the probability density of Lyapunov spectrum collapse to a universal curve, which in turn implies its invariance.

To this goal, we study how the topology is related to extensivity, when inter-connections are modified (in the strength and topology) between subnetworks $G_1$ and $G_2$ forming the multiplex network to maximise the sum of LEs given by
\begin{equation}\label{final_HKS_1_network1}
H_{KS} \approxeq  N\log{(2)} - \epsilon(N_1) S - \gamma l_{12},\nonumber
\end{equation}
where Eq. (\ref{final_HKS_1_network1}) differs from Eq. (\ref{final_HKS_1_network}) in that we consider $S$ instead of $S^{\prime}$, which thus allows us to do the analysis not only for the multiplex network but also for a single layer.

Extensivity in the multiplex network is achieved if $H_{KS}$ is a linear function of $N$, and this leads to the requirement that $\epsilon(N_1) S + \gamma l_{12}= \sigma N$, which using that $\sum_{i=1}^{N_1}\omega_i=\sum_{i=1}^{N_1}d_i\equiv S$ can be written as 
\begin{equation}\label{constrain2}
\epsilon(N_1) \sum_{i=1}^{N_1} \omega_i + \gamma l_{12}= \sigma N +\nu, 
\end{equation}where $\nu$ is a constant.

Equation (\ref{constrain2}) can be further developed to 
\begin{flalign}
\frac{1}{N_1} \sum_{i=1}^{N_1} \left(C_1(\Lambda,N_1) \frac{\omega_{i}}{\omega_{N_1}} + 1 -C_1(\Lambda,N_1) \right) = C_1(\Lambda,N_1) \left(\frac{\bar{d}}{\omega_{N_1}}  - 1 \right) +1=&\nonumber\\ C_1(\Lambda,N_1) [C(\Lambda,N_1)  - 1 ] +1 &= C_2, 
\label{prova2-5}
\end{flalign} 
\noindent
where $\bar{d}=\frac{1}{N_1}\sum_{i=1}^{N_1} \omega_i=\frac{1}{N_1}\sum_{i=1}^{N_1} d_i$ represents the average intra-degree, and 
\begin{equation}
C(\Lambda,N_1)  = \frac{\bar{d}}{\omega_{N_1}}, 
\label{equation-C-1}
\end{equation}
\begin{equation}
C_1(\Lambda,N_1) = \mu_{2N_1-1} =  \epsilon(N_1) \omega_{N_1},  
\label{prova2-1-2} 
\end{equation}
and 
\begin{equation}
\frac{1}{N_1} \sum_{i=1}^{N_1} \left(\epsilon(N_1)\omega_{i} + 2\gamma\alpha \right) = C_2, 
\nonumber
\end{equation}
with $C$ and $C_2$ being constants for the network to be extensive.
 
Equation (\ref{prova2-5}) can also be written for convenienceas  
\begin{equation}
\epsilon(N_1) \bar{d}\frac{N}{2} + \frac{N}{4}(1-\epsilon(N_1) \omega_{N_1})= \sigma N + \nu.
\label{prova2-2-1}
\end{equation}
In the following we analyse one case for multiplex networks and one for a single network, whereas several other cases can be seen in S1 file.

From Eq. (\ref{prova2-1-2}), let us choose $C_1(\Lambda,N_1)=\frac{\omega_{N_1}}{N}$ such that $\epsilon(N_1)=\frac{1}{N}$ and increase the average degree, $\bar{d}$, linearly with $N$ as the network grows, i.e. $\bar{d} = \alpha N + \xi$. Equation (\ref{prova2-2-1}) can be rewritten as  
\begin{equation}
\frac{N}{4} -\frac{\omega_{N_1}}{4} +\frac{\alpha N + \xi}{2} = \sigma N + \nu,  
\nonumber
\end{equation}
and the network will be extensive with $\sigma=\frac{1}{4}-\frac{\omega_{N_1}}{4N}+\frac{\alpha}{2}$. Both $C(\Lambda,N_1) = \frac{\alpha N +\xi}{\omega_{N_1}}$ and $C_2(\Lambda,N_1)$ will not be constant, and therefore no invariant Lyapunov spectrum. The set of LEs in the limit of $N_1 \rightarrow \infty$ is given by $\lambda_{2i} = \log{\left(2 - \frac{1}{N} (\omega_i - \omega_{N_1}) -1\right)}$, i.e. the set of LEs that produce the non-invariant Lyapunov spectrum, and $\lambda_{2i-1} = \log{\left(2 - \frac{\omega_i}{N} \right)}$. Here, it becomes clear that the density of the difference $(\omega_i - \omega_{N_1})$ is crucial for the invariance (or not) of the curve for the Lyapunov spectrum. For a finite $N$, the spectrum of LEs receives a significant contribution from the constant term $\xi$, resulting in an apparent non-invariant curve for the Lyapunov spectrum.

For a single network, if $\epsilon$ in Eq. \eqref{HKS-single} is chosen to make the network extensive, i.e. $H_{KS}\propto N$, it would imply that $\epsilon=\frac{C_1}{\omega_{N_1}}$, with $C_1<1$ being a constant. This choice would also make the curve for the Lyapunov spectrum invariant. 
If the inter-coupling strengths and topology described by $\gamma l_{12}$ of the multiplex network is modified by an evolutionary process that maintains the positiveness of all LEs (which contributes to maintaining extensivity as well), and if $\epsilon$ is chosen to maintain extensivity, i.e. $\epsilon(N_1)=\frac{C_1(N_1)}{\omega_{N_1}}$, where $C_1(N_1)<1$ is a function of the subnetwork topology, then, the invariance of the curve of the Lyapunov spectrum can only be granted if $C(N_1)=\frac{\bar{d}}{\omega_{N_1}}$, where $\bar{d}$ is the average degree and, $C_1(N_1)$ and $C_2(N_1)=C_1(N_1)[C(N_1)-1]+1$ are all invariant with respect to $N$ or $N_1$.

\section{Discussion}
In the theory of information, there are two main quantities involved. The information generated (or lost) and the information exchanged (or shared). This paper deals with the former quantity, whereas the work in \cite{baptista_arxiv2015} is dedicated to the study of the latter quantity. It is worth mentioning that the optimal topologies for the transfer of information found in \cite{baptista_arxiv2015} deviate from the topologies shown to provide extensivity in this work. An extensive network exchanges little information between its layers.

Our analytical and numerical results elucidate the importance of the properties of the structure of complex networks and of the interaction strengths among their constituent parts for extensivity to hold. We show that extensivity can be maintained in an evolving multiplex network by scaling coupling strengths of the dynamics associated to the short- and long-range interactions as to maximise the sum of the positive LEs. For the analytically treated multiplex networks whose both the sum of intra-degrees and sum of inter-degrees scale as $N^{\theta+1},\;\theta>0$, extensivity can be maintained if the intra- and inter-coupling strengths scale as $N^{-\theta}$. Our results for the considered small-world neural networks studied here show that the sum of the positive LEs can be an extensive quantity. It is, of course, an open question whether actual brain networks, which have small-world properties \cite{He01102007}, also evolve by rewiring their connectivity in order to promote an extensive increase in the production of information rate. Since extensivity was achieved here by a non-generic rewiring procedure, one might still conclude that generically an evolving network will not exhibit extensivity.

In order for a network with extensive dynamics to have an invariant curve for the Lyapunov spectrum, it is required that certain network quantities remain invariant during evolution. Even networks evolved by changing their network class, can still be classified as  networks with extensive dynamics and with invariant Lyapunov spectra if these quantities remain invariant during evolution. These quantities are related to the average intra-degree, to the largest eigenvalue of the subnetworks that form the network, and to a ratio of these two quantities.

Concluding, this paper provides a clear message and rigorous results that demonstrate that complex networks do have extensive quantities and therefore, their behaviour can indeed depend on their sizes.

\section{Supporting information}

\paragraph{S1 File.}
\label{S1_file}

\section{Acknowledgments}
We would like to thank Dr N. Rubido for discussions about the extensive character of complex networks, and Dr Rodrigo F. Pereira for valuable discussions, especially with regard to the finite nature of super-extensive behaviour. This work was performed using both the Maxwell high performance computing cluster and the ICSMB cluster of the University of Aberdeen. Both authors acknowledge financial support provided by the EPSRC Ref: EP/I032606/1 grant. C. G. A. contributed to this work while working at the University of Aberdeen and then, when working at the University of Essex.

\end{document}